\documentclass{elsart}
\usepackage{amssymb}

\begin{document}

\begin{frontmatter}

\title{A Simple Method of Calculating Commutators in Hamilton System with Mathematica Software}
\author{Hong-Tao Zhang\thanksref{email}}
\address{
Department of Physics, Peking University, Beijing 100871, P. R.
China }
\date{\today}
\begin{abstract}
As a powerful tool in scientific computation, Mathematica offers
us algebraic computation, but it does not provide functions to
directly calculate commutators in quantum mechanics. Different
from present software packets to deal with noncommutative algebra,
such as NCAlgebra and NCComAlgebra, one simple method of
calculating the commutator in quantum mechanics is put forward and
is demonstrated by an example calculating SO(4) dynamical symmetry
in 3 dimensions Coulomb potential. This method does not need to
develop software packets but rather to directly write program in
Mathematica. It is based on the connection between commutator in
quantum mechanics and Poisson bracket in classical mechanics to
perform calculations. Both the length and the running time of this
example are very short, which demonstrates that this method is
simple and effective in scientific research. Moreover, this method
is used to calculate any commutator in Hamilton system in
principle. In the end some deficiencies and applications are
discussed.
\end{abstract}

\thanks[email]{E-mail address:zhanght123@sina.com}

\end{frontmatter}

\section{INTRODUCTION}
Today scientific computations are classified into two kinds. One
is pure numerical computation, the other is algebraic computation.
The latter treats with symbolic computation rather than simple
numerical computation, thus it holds more general applications in
scientific research. Among much software to implement symbolic
computation, Mathematica, developed by Wolfram Research Inc., is
one of the most popular, influential and active software in
academic and industrial world \cite{math}. It offers us powerful
scientific computational facilities, algebraic computation,
high-quality graphical visualization and user interfaces.
Moreover, its basic system is developed in C program language, and
it is easy to be transplanted into other programs and languages.

Mathematica has been extensively applied in scientific research,
especially in mathematics, physics and chemistry. For instance, it
has been used to solve Sch\"{o}ringer equation for bound states
\cite{luc}, calculate  tree-level multi-particle electroweak
production process \cite{hsi}, and perform numeric computation of
dyadic Green's function \cite{li}. But there still exist many
other problems on symbolic computations to be simplified.
Calculating commutators in quantum mechanics is one of such
problems.

The emergence of the commutator is a fantastic event in physics
\cite{heis}. In classical mechanics there does not exist
noncommutative quantity, while in quantum mechanics, there are two
kinds of quantities: Q quantity and C quantity. Q quantity denotes
noncommutative quantity. For example, when two quantities
$A$, $B$ do not meet commutative law, i.e. $\left[ A,B\right] =AB-BA\neq 0$, we may say that $A$, $%
B$ are two Q quantities. Otherwise, we may say that $A$, $B$ are
two C quantities. In quantum theory, especially in quantum
mechanics, quantum group and statistical physics, we often have to
calculate many commutators. Today the research of the commutator
has initiated noncommutative algebra \cite{noncom1} and
noncommutative geometry \cite{noncom2,noncom3}, which are all hot
topics in physics.

However, in practice, to perform these symbolic calculations is a
tedious and mechanical job for physicists, so it is necessary to
seek a simple and effective method of calculating operators with
computer. It should be noted that even though Mathematica does
provide symbolic computations, it does not include calculating
commutators yet. Therefore, a direct idea is to develop source
codes in Mathematica to compute commutator or anti-commutator.
Along this routine, scientists have developed some software
packets to deal with noncommutative algebra \cite{hel,nug,mak},
such as NCAlgebra \cite{hel} and NCComAlgebra \cite{nug}, to
perform symbolic computations. On one hand, to develop, install
and use these professional packets need many skills and knowledge
of computer, which might be hard work for most physicists. On the
other hand, in these packets, we should set the basic commutative
relations firstly, and then by using computer, we simplify these
operators. The process makes it inconvenient to calculate
commutator with this software. Moreover, some software is
commercial and it would not be used free. Hence, if one simple way
to calculate the commutator is put forward, it will bring us
convenience in practical research. After all, we need a simple
method. Actually, noted that Mathematica provides differential
calculation, and with the relation between commutator in quantum
mechanics and Poisson bracket in classical mechanics, it might be
used to calculate the commutators. Based on this idea, one simple,
easy new method of calculating commutators in Hamilton system is
presented and its process is illustrated by one example in this
article. The method does not need to develop any software packet
or modify any source code in Mathematica and it executes
calculations only with functions in Mathematica software itself.
It should be noted that this method is very simple and it does not
need other extra skills or knowledge of computer. Moreover, this
method can be extended into solving more problems. In the end some
defects and applications of this method are discussed. This method
is expected to simplify many practical computations in further
scientific research.
\section{POISSON BRACKETS AND COMMUTATORS}
Now let us recall the relation between Poisson bracket and
commutator
\begin{equation}
\{A,B\}=\frac 1{i\hbar }[A,B] ,\end{equation} where $[\ ]$ donates
commutator, $\{\}$ donates Poisson bracket\cite{poi}, $i$ is the
imaginary number unit, and $\hbar $ is Planck's  constant. This
relation gives us the connection between the quantum
mechanics and classical mechanics. Assume a Hamilton system $H(p_i,q_i)$, where $%
p_i,q_i$ are canonical coordinates and their conjugate momenta
respectively. For two mechanical quantities $A(p_i,q_i)$,
$B(p_i,q_i)$, the Poisson bracket is defined by \cite{gold}

\begin{equation}
\{A,B\}=\sum_{i}(\frac{\partial A}{\partial q_i}\frac{\partial B%
}{\partial p_i}-\frac{\partial A}{\partial p_i}\frac{\partial B}{\partial q_i%
}).
\end{equation}
From Eq.(1) and (2), we will obtain

\begin{equation}
\lbrack A,B]=i\hbar\sum_{i}(\frac{\partial A}{\partial q_i}%
\frac{\partial B}{\partial p_i}-\frac{\partial A}{\partial p_i}\frac{%
\partial B}{\partial q_i}).
\end{equation}

Eq.(3) provides some clues to  calculate commutators. It should be
noted that the left side of eq.(3) is noncommutative algebra, but
the right-hand side becomes a polynomial in classical mechanics
and there does not exist noncommutative algebra. i.e. eq.(3) give
us a tool to change noncommutative algebra calculations into
commutative algebra calculations. When we compute commutators in
quantum mechanics, we firstly rewrite these commutators into
Poisson brackets in classical mechanics. Then we could use the
functions in Mathematica directly to calculate those Poisson
brackets, and at last we rewrite the results into quantum
operators and get calculating results. Because Mathematica
provides powerful functions, such as Factor, Simplify, D[f,x]
\cite{math}, we can effectively finish complex calculations in a
few minutes. In the next section we will describe this method by
computing SO(4) group in 3 dimensions Coulomb potential.

\section{CALCULATING SO(4)GROUP IN 3 DIMENSIONS COULOMB POTENTIAL WITH MATHEMATICA}

Assume a particle in Coulomb potential $V(r)=-\frac 1r$, we try to
calculate the dynamical symmetry group SO(4). In 3 dimensions
Coulomb potential, there exist three conserved quantities,
Hamiltonian $H$, angular momentum $L$ and Runge-Lenz vector
$R=\frac 12(P\times L-L\times P)-e_r$ \cite{runge}. In order to
know whether angular momentum $L$ and Runge-Lenz $R$ construct
SO(4) Lie group, we should calculate the commutators
$[H,R],[R,R],[R,L],[L,L]$. Firstly we calculate the commutator
$[R,R]$. In fact from the symmetry, we might only calculate the
commutator $[R_x,R_y]$, where $R_x,R_y$ are Runge-Lenz vector $R$
along $x$ and $y$ axis respectively. Generally, calculating these
commutators is a tedious job, but we will show that with
Mathematica it becomes a piece of cake.

First. Simplify the commutator to be calculated with the algebra
relation $P\times L+L\times P=2i\hbar P$. Then, we obtain
\begin{equation}
\lbrack R_x,R_y]=[(P\times L-i\hbar P)_x-e_x,(P\times L-i\hbar
P)_y-e_y].
\end{equation}
In most circumstances, we should simplify the commutators before
we calculate these commutators.

Second. Using eq.(1) and eq.(4), we rewrite above commutator into
Poisson bracket as follows:
\begin{equation}
\lbrack R_x,R_y]=i\hbar \{(P\times L-i\hbar P)_x-e_x,(P\times
L-i\hbar \ P)_y-e_y\},
\end{equation}
here we should note the definitions $(P\times L)_x=i\hbar (l_zp_y-l_yp_z)$, $%
l_x=i\hbar (yp_z-zp_y)$ and $e_x=\frac x{(x^2+y^2+z^2)^{\frac 12}}$, $%
e_y=\frac y{(x^2+y^2+z^2)^{\frac 12}}$ etc.

Third, eq.(5) also can be rewritten into
\begin{eqnarray}
\lbrack R_x,R_y] &=&i\hbar \ {\sum_{i,p_i} }(\frac
\partial {\partial i}((P\times L-i\hbar P)_x-e_x)\frac \partial
{\partial
p_i}((P\times L-i\hbar P)_y-e_y)    \\
&&\ \ -\frac \partial {\partial p_i}((P\times L-i\hbar
P)_x-e_x)\frac
\partial {\partial i}((P\times L-i\hbar P)_y-e_y))  \nonumber,
\end{eqnarray}
where $i=x,y,z$; $p_i=p_x,p_y.p_z$. Using the function in
Mathematica(see Appendix), we could directly calculate the
right-hand of eq.(6) as follows
$\ $%
\begin{equation}
\lbrack R_x,R_y]=-i\hbar (p_yx-p_xy)(P^2-\frac 2r).
\end{equation}
Similarly, we could calculate other commutators with Mathematica.

\begin{equation}
\lbrack H,R_x]=0,
\end{equation}

\begin{equation}
\lbrack L_x,R_y]=-i\hbar (p_xp_zx+p_yp_zy-zp_x^2-zp_y^2-i\hbar
p_z-\frac zr),
\end{equation}

\begin{equation}
\lbrack L_x,L_y]=-i\hbar (xp_y-yp_x).
\end{equation}

Fourth, rewrite above results into the standard quantum mechanics
operators and simplify these results. We get

\begin{equation}
\lbrack R_x,R_y]=-2i\hbar HL_z,
\end{equation}

\begin{equation}
\lbrack H,R]=0,
\end{equation}

\begin{equation}
\lbrack L_x,R_y]=i\hbar R_z,
\end{equation}

\begin{equation}
\lbrack L_x,L_y]=i\hbar L_z.
\end{equation}

Up to now we have finished calculating such tedious commutators.
From above relations, we can see that the Rungle-Lenz vector $R$
and angular momentum $L$ construct the SO(4) dynamical symmetry,
which is higher than geometry symmetry SO(3). This is a well-known
result in quantum mechanics \cite{pauli,schiff}. What we want to
do here is not to prove or check this result, but to demonstrate
that our method is efficient. It should be noted that the program
is very short(only 20 lines) and the running time is only half a
minute (PC 300, 64MB RAM)! In contrast to calculating above
commutators by hand, it is faster and more exact. The program and
some expeditions are listed in Appendix and there the function
Commutator3D can be used to compute any commutator in 3 dimensions
Hamilton system.

\section{DISCUSSIONS}

Because the connection between the commutator and Poisson bracket
only helps to calculate the Hamilton system in quantum mechanics,
our method might be unsuitable for the second quantize system.
However, it does provide a simple and powerful method of
calculating or checking commutators on dynamics symmetry and Lie
algebra in quantum mechanics and classical mechanics, especially
when the Hamiltonian contains complicated potentials. In
principle, this method is used to calculate any commutator in
Hamilton system. We test our method in several problems, such as
calculating Lie algebra or Lie group in isotropic harmonic
potential \cite{hill}, screened Coulomb potential and screened
isotropic harmonic potential \cite{wu}. All of the results have
demonstrated that this method is simple, effective and robust. We
expect this method could help physicists solve such problems
easily. Moreover, this method is easy to be applied into other
mathematical software, such as Maple, Matlab. Meanwhile our method
indeed shows that for calculating most commutators in Hamilton
system, Mathematica could solve these problems by itself.

Of course, it is interesting and valuable to develop these
software packets \cite{hel,nug,mak} to perform symbolic
calculations, because they will release us from the bondage of
mechanical and tedious work, and help us conduct real scientific
research. In contrast to using these software, our method seems to
be simpler and easier for most physicists. Of course, this method
could not take the place of these software packets and itself is
only supplement to these software packets. Actually we also hope
that our method could give some clues in developing noncommutative
algebra software and make it more powerful.

\section*{ACLKNOWLEDGEMENTS}

We thank Mis. Ge Hua for giving help in revising manuscript.

\section*{APPENDIX}
(*This program is used to calculate the SO(4) group in 3D Coulomb
potential \ V(r) = -1/r *)\\
(* Comumutator3D is a function to calculate commutators in 3
dimensions, where f and g are variables to be computed *)
\\
(* p$_i$ and q$_i$ (i=1,2,3)  are canonical coordinates and their
conjugate momenta respectively *)
\\
(*ham is Hamiltonian of system, h is Planck's constant, l is
angular momentum and r is Rungz-Lenz  vector*)
\\
\\
Commutator3D[f\_, g\_]:= i h \{D[f, q$_1$] D[g, p$_1$]- D[f,
p$_1$] D[g, q$_1$]+D[f, q$_2$] D[g, p$_2$] -D[f, p$_2$]
D[g,q$_2$]+ D[f, q$_3$] D[g, p$_3$]- D[f, p$_3$] D[g, q$_3$] \}\\
\\
ham=\{(p$_1$\symbol{94}2+p$_2$\symbol{94}2+p$_3$\symbol{94}2)/2\}-1/Sqrt[q$_1$\symbol{94}2+q$_2$\symbol{94}2+q$_3$\symbol{94}2)]
\\
\\
l$_1$=q$_2$p$_3$-q$_3$p$_2$\\ l$_2$=q$_3$p$_1$-q$_1$p$_3$\\
l$_3$=q$_1$p$_2$-q$_2$p$_1$\\
\\
r$_1$=l$_3$p$_2$-l$_2$p$_3$-i h
p$_1$-q$_1$/Sqrt[q$_1$\symbol{94}2+q$_2$\symbol{94}2+q$_3$\symbol{94}2)]\\
r$_2$=l$_1$p$_3$-l$_3$p$_1$-i h
p$_2$-q$_2$/Sqrt[q$_1$\symbol{94}2+q$_2$\symbol{94}2+q$_3$\symbol{94}2)]\\
r$_3$=l$_2$p$_1$-l$_1$p$_2$-i h
p$_3$-q$_3$/Sqrt[q$_1$\symbol{94}2+q$_2$\symbol{94}2+q$_3$\symbol{94}2)]\\
\\
s$_1$=Commutator3D[r$_1$,r$_2$]\\s$_2$=Commutator3D[r$_1$,ham]
\\s$_3$=Commutator3D[l$_1$,r$_2$]\\s$_4$=Commutator3D[l$_1$,l$_2$]\\
f$_1$=Simply[s$_1$]\\ f$_2$=Simply[s$_2$]\\ f$_3$=Simply[s$_3$]\\
f$_4$=Simply[s$_4$]

\end{document}